\documentclass[%
 reprint,
superscriptaddress,
 amsmath,amssymb,
 aps,
 prl,
]{revtex4-2}

\usepackage{graphicx}
\usepackage{dcolumn}
\usepackage{bm}
\usepackage{subcaption,amsmath,gensymb,esvect,xcolor,xr,multirow,commath}
\usepackage[colorlinks=true,citecolor=blue]{hyperref}
\usepackage[normalem]{ulem}

\begin{document}

\preprint{APS/123-QED}

\title{Giant Valley-Polarized Spin Splittings in Magnetized Janus Pt Dichalcogenides}

\author{Shahid Sattar}
 \email{shahid.sattar@lnu.se}
\affiliation{Department of Physics and Electrical Engineering, Linnaeus University, Kalmar SE-39231, Sweden}
 \author{J. Andreas Larsson}
 \affiliation{Applied Physics, Department of Engineering Sciences and Mathematics, Lule\aa\,University of Technology, Lule\aa\, SE-97187, Sweden}
 \author{C. M. Canali}
\affiliation{Department of Physics and Electrical Engineering, Linnaeus University, Kalmar SE-39231, Sweden}
   \author{Stephan Roche}
\affiliation{Catalan Institute of Nanoscience and Nanotechnology (ICN2), CSIC and BIST, Campus UAB, Bellaterra, 08193 Barcelona, Spain}
\affiliation{ICREA, 08101 Barcelona, Spain}
\author{Jose H. Garcia}
 \email{josehugo.garcia@icn2.cat}
 \affiliation{Catalan Institute of Nanoscience and Nanotechnology (ICN2), CSIC and BIST, Campus UAB, Bellaterra, 08193 Barcelona, Spain}
\date{\today}

\begin{abstract}
We reveal giant proximity-induced magnetism and valley-polarization effects in Janus Pt dichalcogenides (such as SPtSe), when bound to the Europium oxide (EuO) substrate. Using first-principles simulations, it is surprisingly found that the charge redistribution, resulting from proximity with EuO, leads to the formation of two K and K$^{'}$valleys in the conduction bands. Each of these valleys displays its own spin polarization and a specific spin-texture dictated by broken inversion and time-reversal symmetries, and valley-exchange and Rashba splittings as large as hundreds of meV. This provides a platform for exploring novel spin-valley physics in low-dimensional semiconductors, with potential spin transport mechanisms such as spin-orbit torques much more resilient to disorder and temperature effects. 
\end{abstract}

\maketitle

One of the key aspects of van der Waals (vdW) heterostructures is the proximity effect, where  interlayer hybridization combines with the electronic properties of their individual constituents to produce synergistic behaviors \cite{NovoselovScience2016,GarciaChemSocRev2018,huang2020emergent,sierra2021van}. A salient illustration of such emerging properties occurs in graphene-based heterostructures, since graphene has exceptional transport properties due to its linear dispersion at two nonequivalent valleys in the Brillouin zone, but lacks spin functionality owing to its very small intrinsic spin-orbit coupling (SOC). However, it  displays gate-tunable spin-dependent phenomena at room temperature \cite{ghiasi2017large,benitez2018strongly,BenitezNatMat2020,khokhriakov2020gate} and spin-valley coupling \cite{cummings2017giant,Dyrda2017,Gmitra2017} when combined with high SOC materials. Graphene also displays proximity-induced magnetism which is evidenced by a plethora of interesting gate-dependent magnetoresitive phenomena \cite{yang2013proximity,mendes2015spin,wang2015proximity,wei2016strong,karpiak2019magnetic,tang2020magnetic,GhiasiNatNano2021}. The incorporation of recently discovered materials with exotic individual properties into vdW heterostructures is a unique way to discover and engineer disruptive material functionalities, and to efficiently realize complex effects, such as, valley polarization, SOC-exchange swapping \cite{zhang2016large,seyler2018valley,norden2019giant,Zollner2020d}, and symmetry-enhanced spin-orbit torques (SOT) \cite{MacNeill2017,PRLJohansen2019,Dolui2020,Sousa2020}, where the combination of magnetism, SOC and valley control are required. 

The recent synthesis of low-symmetry materials such as Fe$_3$GeTe$_2$ \cite{Deng2018}, 1T'-WTe$_2$ \cite{Tang2017} and Janus SMoSe \cite{Lu2017}, has shown new ways to control the spin degree of freedom by removing symmetry constraints and led to the observation of unusual phenomenon such as canted quantum spin Hall effect \cite{Garcia2020}, persistent spin-textures \cite{Vila2020}, multi-directional spin Hall effect \cite{Safeer2019e,Song2020,Zhao2020}, and interface-free Rashba effect \cite{Cheng2013b}. However, these systems are typically metastable \cite{DuerlooNatCom2014,VoiryCSR2015a}, a challenge that can be overcome by encapsulation or via the substrates, and that also open a door for novel collective effects. Transition metal dichalclogenides (TMDs) in their Janus configuration (where one of the dichalcogenides of the TMD is replaced by a different one) \cite{Cheng2013b} are interesting low-symmetry materials since they possess an intrinsically broken inversion symmetry that yields to significant Rashba splittings and spin-momentum locking, and then offering SOT mechanisms \cite{Cheng2013b,tao2019electronic}. These two features are essential elements for realizing spin-orbit torques (SOTs) \cite{Miron2011,RMPManchon_2019}, a phenomenon where the magnetization direction of a magnetic system is electrically-controlled via the SOC, and that is very promising for efficient nonvolatile memories with nanosecond dynamics \cite{Garello2014a,Manipatruni2019}.   

Ultrathin Pt dichalcogenides (PtX$_2$ (X=S, Se, Te)) constitute a particularly interesting class of materials with a very strong and highly tunable SOC that bestow it with exotic electronic and spin properties \cite{WangNanoLett2015,Li2016a,yao2017direct,ciarrocchi2018thickness,villaos2019thickness}. Their most stable geometrical phase is the 1T-phase (P$\bar{3}$m$1$) which displays a semiconducting behavior \cite{WangNanoLett2015}. Monolayer (1L) SPtSe was recently synthesized in a Janus form \cite{sant2020synthesis}, enabling strong spin-momentum locking due to their low-symmetry nature \cite{yao2017direct}. To fruitfully exploit such spin-momentum locking for SOT applications, it is imperative to determine the spin physics of the Janus-TMD coupled to a magnetic material. Such a coupling should be sufficiently strong to allow for the magnetization control via the electronic spins propagating inside the TMD, but weak enough to at least partially preserve the spin-momentum locking \cite{RMPManchon_2019}. In Pt-based TMDs, some studies suggest the formation of magnetism \cite{avsar2019defect,avsar2020probing}, but to date the consequences of proximity effects between platinum-based TMDs and magnetic substrates have not yet been explored. 

In this letter, we predict that low-symmetry Janus SPtSe on EuO is a suitable material combination that generates large exchange coupling and valley-polarization, while preserving spin-momentum locking. Specifically, the EuO substrate induces the formation of two time-reversal related valleys with C$_{3v}$ symmetry at low symmetry points along the $\Gamma$-$K$/$\Gamma$-$K'$ paths. These valleys display a {\it large spin-splitting of several hundreds meV}, which are traced back to the joint contributions of Zeeman-like and Rashba interactions. We also discuss how the combination of these effects could lead to {\it a giant spin-orbit torque and a current-driven magnetic anisotropy}, originating from the valley's special point group symmetry. {Since these materials possess a modest lattice mismatch amenable to epitaxial growth \color{red}and a low growth temperature similar to other recently synthesize systems} \cite{Mallick2020,Rosenberger2021}, we propose them as the first feasible 2D material with all ingredients required for efficient spin-orbit torques and magnetic-field-free switching of the magnetization.

\begin{figure}[!t]
\includegraphics[width=0.5\textwidth]{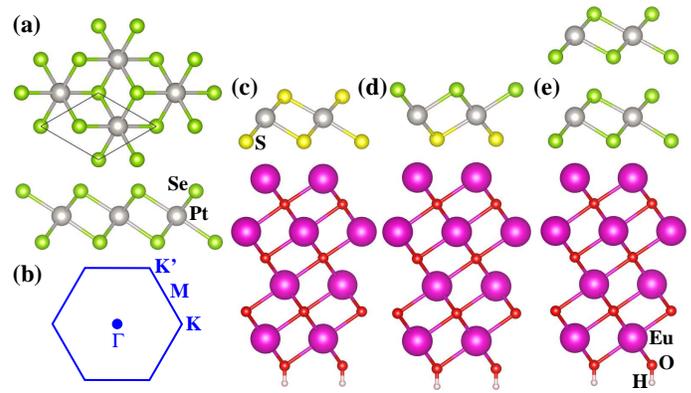}
\caption{(a) Top and side views of 1L PtSe$_2$ (black rectangle shows unit cell). (b) Hexagonal Brillouin zone with high-symmetry points. Minimum energy configurations of (c) PtS$_2$, (d) Janus SPtSe, and (e) 2L PtSe$_2$ on EuO substrate.}
\label{fig:fig1}
\end{figure}

To start with, we aim at clarifying if the strong SOC and broken inversion symmetry in SPtSe can lead to a large spin-splitting and spin-momentum locking. Besides, one key question is whether a magnetic insulating substrate such as EuO can induce a strong exchange coupling in SPtSe which will be essential for SOT applications. To answer those questions, we performed fully-relativistic first-principles calculations using the Vienna Ab-initio Simulation Package (VASP) \cite{vasp} to determine the geometrical, electronic, and spin properties of the Pt-dichalcogenides in the proximity to the magnetic substrate EuO. Technical details of the simulations are presented in the Supplementary Information \cite{suppmat}. We found that all freestanding TMDs are stable in the configuration shown in Fig. \ref{fig:fig1}.a. Both PtSe$_2$ and PtS$_2$ have similar electronic structures displaying a semiconducting behavior with an indirect band-gap and no band-splitting \cite{suppmat}. The lack of spin-splitting, despite the presence of large SOC, is the result of time-reversal-symmetry Kramer's degeneracy combined with  inversion symmetry (one of the symmetries of the lattice $\bar{3}$m point group), which leads to two-fold degenerate bands over the whole BZ. It is pertinent to mention that the calculated band-gaps values are underestimated; for example, 1L PtSe$_2$ displays an indirect band-gap of 1.20 eV which is significantly smaller than the experimentally reported value of 2.2$\pm$0.1 eV \cite{ciarrocchi2018thickness}. Therefore, an energy shift of $\sim$0.9 eV in the conduction band is needed to compensate and match the experimental values \cite{sajjad2018strongly}. We also highlight the lack of valleys at the $K$ and $K'$ points in the BZ defined in Fig. \ref{fig:fig1}.b, which is a significant difference between these structures and the more commonly found TMDs in the 2H-phase. 

In the Janus TMD SePtS, the presence of dissimilar chalcogen atoms breaks inversion symmetry by removing a vertical mirror plane reducing the point group to 3m; this lifts the band degeneracy with a significant splitting in the range of 10-100 meV \cite{suppmat}. An analysis of different kinds of stacking showed that the rock-salt crystal structure of EuO oriented along the (111) hexagonal face, sketched in Fig. \ref{fig:fig1}.c, is the ideal orientation for epitaxial growth, since in this case the Janus SPtSe presents a very small amount of strain of 0.3$\%$, while all the other symmetric structures have strains of about $3\%$ \cite{suppmat}. After the relaxation procedure, we determined that Pt-atoms prefer to sit on top of the Eu-atoms, resembling the MoTe$_2$/EuO structure \cite{zhang2016large}. We found that EuO has a ferromagnetic ground state, and a lattice constant of 3.65 {\AA} on its (111) hexagonal face. To identify the proximity-induce exchange interaction, we have performed calculations in the absence of SOC and found the typical exchange interactions to be very large, $\sim 400$ meV \cite{suppmat}. We also observed the unexpected formation of valleys in the vicinity of high-symmetry $K$ and $K$' point of the BZ. To our knowledge, this is the first example of proximity-induced valley formation in 2D materials. Since Eu and Pt are very heavy elements, we expect strong possible SOC effects at these induced $K$-points that could be used for purely electrical manipulation of the magnetic moments in EuO. Systems displaying valleys at the $K$-points with 3m point group symmetries are very important for SOT since they allow for optimal Rashba-Edelstein effect and efficient SOT \cite{Offidani2017,Sousa2020} due to the possibility of vertical spin-splitting and symmetric spin-momentum locking in the vicinity of two time-reversal symmetric $K$-points, which also enables a novel kind of anisotropy-like SOT \cite{PRLJohansen2019} hitherto only observed in Fe$_3$GeTe$_2$ \cite{Alghamdi2019}. {\color{red} However, unlike Fe$_3$GeTe$_2$, Pt-based TMDs are semiconductors, which allows for the possibility of gate-controlled SOTs, while the large Zeeman-exchange and SOC open the door for very large efficiency even with low-current densities.}

\begin{figure}[!t]
\includegraphics[width=0.5\textwidth]{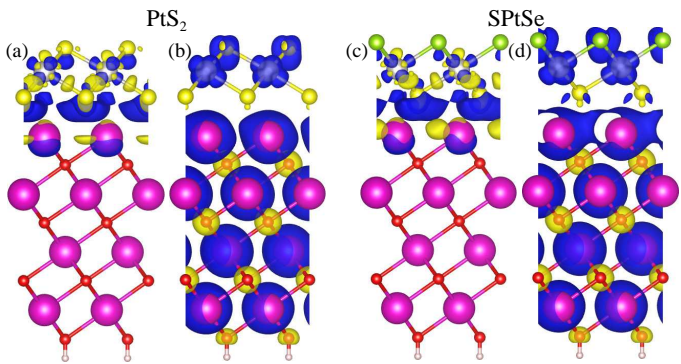}
\caption{(a,c) Charge density difference (isovalue: $2.5\times10^{-3}$ electrons/bohr$^{3}$), and (b,d) Spin density (isovalue: $3.0\times10^{-3}$ electrons/bohr$^{3}$ ) of 1L PtS$_2$ and Janus SPtSe on EuO substrate, respectively. In (a,c), blue and yellow colors represent charge accumulation and depletion, wheres in (\textcolor{red}{b},d), blue and yellow colors represent spin-up and spin-down densities, respectively.}
\label{fig:fig3}
\end{figure}

The origin of the large proximity-induced magnetism is actually understood by the observed significant charge redistribution between the proximitized materials and the surface of the EuO substrate. Fig. \ref{fig:fig3}.a shows the charge density difference which evidences a substantial charge accumulation on the bottom chalcogen atom compared to the top one with charge-density extending inside the van der Waals gap. By further performing a Bader charge analysis, one demonstrates that bottom S atom acquires 0.52 $|e|$ charge from the surrounding atoms (Eu loses 0.49 $|e|$ and Pt loses 0.06 $|e|$) resulting in $n$-doping of the system. Fig. \ref{fig:fig3}.b clarifies that such a charge redistribution is accompanied by a transfer of spin-polarized electrons from Eu to Pt atoms, which leads to the formation of a magnetic moment of magnitude  $+0.2\mu_B$ located on the Pt $d$-orbitals. We also observe S atoms to be anti-ferromagnetically coupled to it by having magnetic moments of $+0.1 \mu_B$ and $-0.02 \mu_B$ located on $p$-orbitals for the top and bottom S atoms, respectively. In other words, non-magnetic few-layer PtX$_2$ (X=S, Se) becomes magnetic on the EuO substrate due to the unique anti-ferromagnetic coupling of the S atoms. Looking at the EuO substrate itself, Eu and O atoms are also anti-ferromagnetically coupled to each other having a total magnetic moment of $+7\mu_B$ and $-0.1\mu_B$ sitting on $f$- and $p$-orbitals, respectively. The system also become $n$-doped due the charge transfer between the chalcogen and surrounding atoms. These results are further validated by the atom-projected density of states that shows a strong orbital hybridization between Pt(d), S(p) and Eu(d) orbitals \cite{suppmat}. Since it is quite difficult to grow monolayer systems in general, and the electronic properties of PtSe$_2$ are remarkably dependent on the crystalline structure \cite{WangNanoLett2015,Li2016a,yao2017direct,ciarrocchi2018thickness,villaos2019thickness}, we also  evaluated the bilayer case in the AA-stacking.

\begin{figure}[!t]
\includegraphics[width=0.5\textwidth]{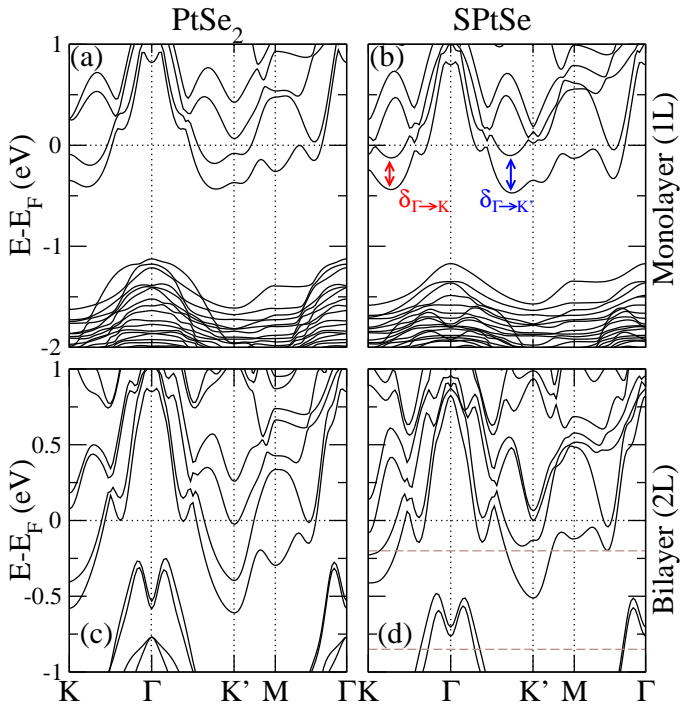}
\caption{(a,c) Electronic band structure of 1L and 2L PtSe$_2$ and (b,d) Janus SPtSe on EuO substrate, respectively. The SOC effects are included for each case.}
\label{fig:fig2}
\end{figure}

We now discuss the results of the calculations when both magnetism and SOC are included, for the monolayer and bilayer cases of  PtSe$_2$ and SPtSe on the EuO substrate. The binding energy and interlayer distance for these calculations are given in Table \ref{table:table1}. The binding energy is calculated using the definition: E$_b=E({\rm PtSe_2/\rm EuO})-E({\rm PtSe_2})-E({\rm EuO}),$ where $E({\rm PtSe_2/\rm EuO})$ is the total energy of the combined system, $E({\rm PtSe_2})$ is the total energy of detached PtSe$_2$, and $E({\rm EuO})$ is the total energy of the EuO substrate. The large value of E$_b$= -1.70 eV (-1.84 eV for 2L PtSe$_2$ on EuO) confirms the presence of strong interlayer interactions. We note that the binding energy is even larger for the lower atomic-size chalcogen PtS$_2$ relatives of PtSe$_2$ with lesser interlayer distance,  providing a stronger attachment of the former to EuO substrate albeit perturbing the electronic dispersion as discussed below. These results are consistent with experimental data on Pt (111) substrate where an interlayer distance of 2.28 \AA\,is observed \cite{WangNanoLett2015}. In Fig. \ref{fig:fig2}.(a-d) the band structures for all the three cases are reported, and unveil a novel effect not present in previous calculations, namely a valley-dependent band splittings highlighted in Fig. \ref{fig:fig2}.b for 1L Janus SPtSe. Such an effect provides convincing evidence of the simultaneous action of exchange and SOC interactions. We estimate the (Zeeman) exchange strength by computing the average of the splitting at each valley $\Delta_{\rm ex} \approx (\delta_{\Gamma\rightarrow K}+\delta_{\Gamma\rightarrow K'})/2$, which leads to values between 190 meV to 360 meV as reported in Table \ref{table:table1}. Valley polarizations are actually quite typical of magnetized TMDs in their 1H structural phase, being a consequence of the hexagonal symmetry (D$_{3h}$ point group) and broken inversion symmetry that create a specific SOC term named valley-Zeeman, which behaves as an effective Zeeman field opposite sign at each valley. Therefore, the combined effect of this valley-Zeeman interaction with a true Zeeman-field yields a valley-dependent Zeeman splitting with a splitting strength piloted by the SOC. We note that in principle, Pt-based TMDs are most stable in the tetragonal phase (D$_{3d}$ point group), a phase preserving inversion symmetry, which forbids such a type of valley-Zeeman SOC term. However, the presence of the substrate breaks such a symmetry whereas the preserved three-fold rotational symmetry induces the formation valley-Zeeman SOC terms at the $K$ points. Consequently, the valley-polarization predicted by our calculations demonstrate  large proximity-induced magnetism and lifting of the inversion symmetry in these Pt-based TMD supported by a EuO substrate.

\begin{table}[!b]
\centering
\caption{Binding energy (E$_\text{b}$), interlayer distance (d), conduction band spin-splittings along the $\Gamma$$\rightarrow$K ($\delta_{\text{ $\Gamma$$\rightarrow$K}}$) and $\Gamma$$\rightarrow$K' ($\delta_{\text{ $\Gamma$$\rightarrow$K'}}$) paths, Zeeman exchange strength ($\Delta_{\rm ex}$) and average difference in splittings ($\lambda_{vz}$). }
\begin{tabular}{|c|c|c|c|c|c|c|c|c} 
\hline\hline
                       &    & \begin{tabular}[c]{@{}c@{}}E$_{\text{b}}$ \\(eV)\end{tabular}& \begin{tabular}[c]{@{}c@{}}d\\(\AA)\end{tabular} & \begin{tabular}[c]{@{}c@{}}$\delta_{\text{ $\Gamma$$\rightarrow$K}}$ \\(meV)\end{tabular} & \begin{tabular}[c]{@{}c@{}}$\delta_{\text{ $\Gamma$$\rightarrow$K'}}$ \\(meV)\end{tabular} & \begin{tabular}[c]{@{}c@{}}$\Delta_{\rm ex}$\\(meV)\end{tabular} & \begin{tabular}[c]{@{}c@{}}$\lambda_{vz}$\\(meV)\end{tabular} \\ 
\hline
\multirow{2}{*}{PtS$_2$}  & 1L & -1.94 & 2.05 & 362  & 355 & 358.5 &3.5  \\ 
\cline{2-8}
                          & 2L & -2.05 & 2.04 & 331  & 270 & 300.5 &30.5  \\ 
\hline
\multirow{2}{*}{SPtSe}    & 1L & -2.04 & 2.07 & 313  & 372 & 342.5 &29.5  \\ 
\cline{2-8}
                          & 2L & -1.87 & 2.05 & 185  & 345 & 265 &80  \\ 
\hline
\multirow{2}{*}{PtSe$_2$} & 1L & -1.70 & 2.20 & 215  & 253 & 234 &19  \\ 
\cline{2-8}
                          & 2L & -1.84 & 2.18 & 171  & 214 & 192.5 &21.5 \\ 
\hline
\end{tabular}
\label{table:table1}
\end{table}

The strength of the valley-Zeeman interaction is estimated by computing the average difference in the splitting at each valley $\lambda_{\rm vz} \approx \abs{\delta_{\Gamma\rightarrow K}-\delta_{\Gamma\rightarrow K'}}/2$. The values in Table \ref{table:table1} show Zeeman interactions ranging from 3 meV to 80 meV in the symmetric structures. These values are extraordinarily large compared to other conventional TMDs, as a consequence of the strong SOC introduced by the heavy metals. We however note that such estimation does not include possible additional contribution from Rashba SOC stemming from the interaction with the substrate, which should slightly renormalize the band splitting. 

\begin{figure}[!htb]
\includegraphics[width=0.5\textwidth]{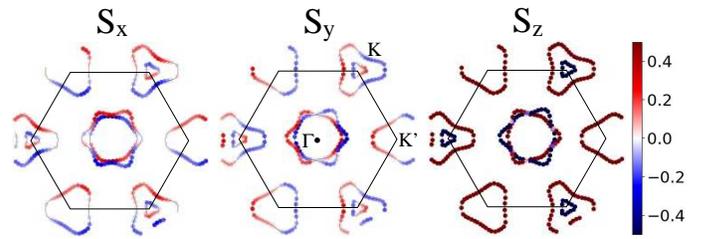}
\caption{2D spin-textures of Janus 2L SPtSe on EuO substrate. Refer to brown lines in Fig. \ref{fig:fig2}.d, inner contours corresponds to valence bands passing through an energy $E=-0.85$ eV whereas the outer contours (corners) belongs to conduction bands passing through an energy $E=-0.20$ eV, respectively. \textcolor{red}{The red/blue colors indicate spin up/down.}}
\label{fig:fig4}
\end{figure}

We finally discuss the results obtained for the asymmetric or Janus structure. In Table \ref{table:table1} a substantial valley polarization is observed, that is 160 meV and 60 meV for the bilayer and monolayer structures respectively. This is three to ten times larger than their symmetric counterparts. Since the structures are formed by the same atoms, such a change can be attributed to the giant Rashba effect expected in this phase. \textcolor{red}{It is pertinent to mention that Janus SPtSe on EuO show these effects regardless of the presence of S or Se adjacent to the substrate \cite{suppmat}}. It is complicated to extract the Rashba splitting since its contribution is mixed with different interactions. Therefore, we assume no Rashba interaction in the symmetric structure and further subtract the average spin polarization for the Janus case. Following this procedure, we evaluate a Rashba coupling ranging between 30 meV and 60 meV, which is again much larger than any other studied Janus-like system to date. 

To determine whether the systems in indeed subject to a strong Rashba SOC, the spin-textures are computed using a 2D k-mesh ($k_x\times k_y:15\times15$) centered at the $\Gamma$-point ($k_z=0$) and are presented for the energy contours $E=-0.20$ eV and $E=-0.85$ eV in Fig. \ref{fig:fig4}. The colors indicates opposite spin-polarities along the projected $x$, $y$, and $z$ directions, represented also in the labels above. In the valence bands, around the $\Gamma$ point, we see a close contour which points to a mexican-hat shaped dispersion, although it is tilted due to the effect of the exchange interaction. This contour displays spin-momentum locking which is opposite for the two concentric bands and hints at potentially observing large spin-orbit torque in experiments. However, in the conduction bands, the contour crosses the $K$/$K'$ points showing two distinct features: (i) strong spin-momentum locking in isolated bands, (ii) an energy contour with $C_{3v}$ point group symmetry. The first feature allows optimal spin-to-charge conversion via the Rashba-Edelstein effect which is a precursor of SOT \cite{Offidani2017,Sousa2020}. The out-of-plane spin component reinforces the idea of a concomitant existence of exchange and Rashba interactions. The presence of a Fermi-contour with three-fold rotational invariance is a necessary condition for a novel kind of anisotropy-like SOT \cite{PRLJohansen2019} hitherto observed only in Fe$_3$GeTe$_2$ \cite{Alghamdi2019}.

In conclusion, we employed first-principles calculations to investigate proximity effects in ultrathin Pt dichalcogenides and Janus SPtSe on magnetic EuO substrate. Substantial charge redistribution \textcolor{red}{was} found within these systems, \textcolor{red}{resulting in shifting} CBMs in the vicinity of high-symmetry K and K$^{'}$ points, \textcolor{red}{with the formation of} multiple valleys. \textcolor{red}{Very importantly, the broken inversion and time-reversal symmetries, together with  proximity effects from the magnetic substrate, generate huge spin-splittings (of the order of several hundred meV)  in the conduction band of few-layer PtX$_2$ (X=S, Se) and Janus SPtSe.} Moreover, these ultrathin systems become magnetic, hosting magnetic moments at different atomic sites with an anti-\textcolor{red}{ferromagnetic} coupling between opposite S atoms demonstrating spin-valley polarization. These findings provide a versatile platform to explore spin-valley physics in Pt dichalcogenides and leads to their potential electronic and spintronics applications. { \color{red} Nevertheless, efforts are still necessary  to improve the air-sensititvy of the Janus structure and increase the  curie temperature of EuO}

Fruitful discussions with Udo Schwingenschl\"{o}gl and Muhammad Tahir are greatly acknowledged. We thank Knut och Alice Wallenberg foundation, Kempestiftelserna and Carl Tryggers Stiftelsen for financial support. The computations were enabled by resources provided by the Swedish National Infrastructure for Computing (SNIC) at HPC2N and NSC partially funded by the Swedish Research Council through grant agreement no. 2018-05973. ICN2 authors were supported by \textcolor{red}{King Abdullah University
of Science and Technology (KAUST) through the award OSR-2018-CRG7-3717 from the Office of Sponsored Research (OSR) and} by the European Union Horizon 2020 research and innovation program under Grant Agreement No. 881603 (Graphene Flagship). ICN2 is funded by the CERCA Programme/Generalitat de Catalunya, and is supported by the Severo Ochoa program from Spanish MINECO (Grant No. SEV-2017-0706 and MAT2016-75952-R).

\bibliographystyle{apsrev4-1}
\bibliography{main}

\end{document}